\documentclass[11pt]{article}
 
\usepackage[T1]{fontenc}
\usepackage[utf8]{inputenc}
\usepackage{mathpazo}              
\usepackage[scaled=0.9]{helvet}    
\usepackage{microtype}             
\usepackage{hyphenat}              
 \usepackage{tabularx}
\usepackage{geometry}
\usepackage{setspace}
\usepackage{tikz}
\usepackage{eso-pic}
\usetikzlibrary{calc}

\AddToShipoutPictureBG*{%
  \AtPageUpperLeft{%
    \begin{tikzpicture}[remember picture, overlay]
      \node[anchor=north west, inner sep=0] (logo)
        at ($(current page.north west)+(0.85in,-0.35cm)$)
        {\includegraphics[height=1.69cm]{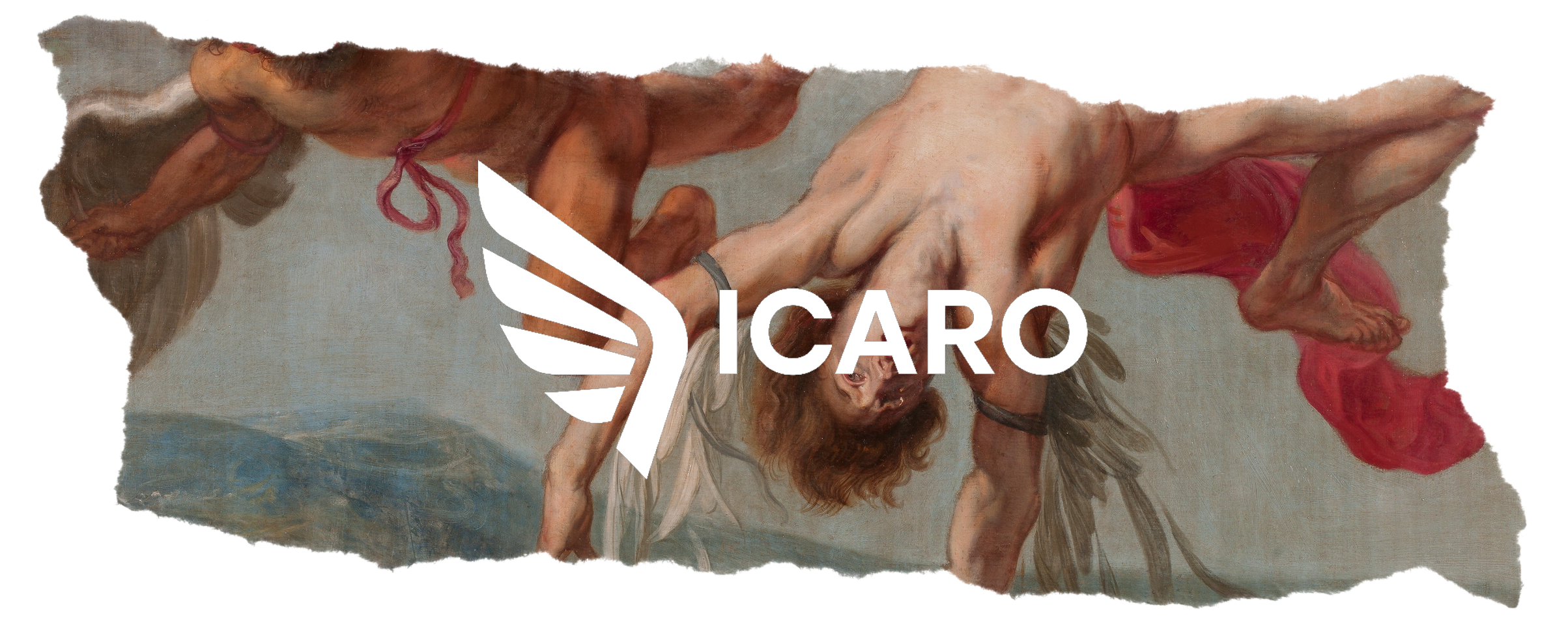}};

      \node[anchor=north east, font=\normalsize]
        (date) at ($(current page.north east)+(-0.85in,-1.2cm)$)
        {\today};

      \draw[line width=0.4pt, black!60]
        ($(current page.north west)+(0.85in,-2.05cm)$) --
        ($(current page.north east)+(-0.85in,-2.05cm)$);
    \end{tikzpicture}%
  }%
}
\usepackage{authblk}
\usepackage{titlesec}
\titleformat{\section}{\large\bfseries}{\thesection.}{0.5em}{}
\titleformat{\subsection}{\normalsize\bfseries}{\thesubsection.}{0.5em}{}
\titleformat{\paragraph}[runin]{\normalsize\bfseries}{}{0em}{}[.]
\titlespacing*{\section}{0pt}{1.5em}{0.5em}
\titlespacing*{\subsection}{0pt}{1.2em}{0.4em}
\titlespacing*{\paragraph}{0pt}{0.8em}{0.5em}
 
\usepackage{booktabs}              
\usepackage{array}
 
\usepackage{graphicx}
\usepackage[font=small, labelfont=bf, labelsep=period]{caption}
 
\usepackage{enumitem}
\setlist{nosep, leftmargin=1.5em}  
 
\usepackage[colorlinks=true, citecolor=black!70!blue, linkcolor=black!70!blue, urlcolor=black!70!blue]{hyperref}
\usepackage{csquotes}
 
\renewenvironment{abstract}{%
  \small
  \begin{center}
    \bfseries Abstract
  \end{center}
  \par\noindent\ignorespaces
}{\par\bigskip}
 
\title{\vspace{-1em}\textbf{Agentic Microphysics:\\[0.3em] A Manifesto for Generative AI Safety}\vspace{0.5em}}
 
\author[1,3]{F. Pierucci}
\author[1,2]{M. Prandi}
\author[1,4]{M. Bracale Syrnikov}
\author[1,2]{M. Galisai}
\author[1,2]{P. Bisconti Lucidi}
 
\affil[1]{\small DEXAI -- Icaro Lab}
\affil[2]{\small Sapienza University of Rome}
\affil[3]{\small Sant'Anna School of Advanced Studies}
\affil[4]{\small VU Amsterdam}
\date{}
 
\begin{document}
 
\maketitle
\thispagestyle{plain}
 
\begin{abstract}
This paper advances a methodological proposal for safety research in agentic AI. As systems acquire planning, memory, tool use, persistent identity, and sustained interaction, safety can no longer be analysed primarily at the level of the isolated model. Population-level risks arise from structured interaction among agents, through processes of communication, observation, and mutual influence that shape collective behaviour over time. As the object of analysis shifts, a methodological gap emerges. Approaches focused either on single agents or on aggregate outcomes do not identify the interaction-level mechanisms that generate collective risks or the design variables that control them. A framework is required that links local interaction structure to population-level dynamics in a causally explicit way, allowing both explanation and intervention. We introduce two linked concepts. \emph{Agentic microphysics} defines the level of analysis: local interaction dynamics where one agent's output becomes another's input under specific protocol conditions. \emph{Generative safety} defines the methodology: growing phenomena and elicit risks from micro-level conditions to identify sufficient mechanisms, detect thresholds, and design effective interventions.
\end{abstract}
 
\section{Introduction}
 
Safety research has largely treated the individual model as its primary object of analysis. A system is evaluated for its outputs, its alignment, or its tendency to produce harmful content under prompt-based interaction. That framing was broadly adequate when the dominant deployment pattern involved a single model answering a single query. It becomes insufficient once systems acquire architectural features associated with agency, including planning, memory, tool use, persistent identity, and extended interaction \cite{Guo2024,Tran2025}. Under those conditions, the relevant object of analysis shifts from the isolated model to populations of interacting agents.
 
Multi-agent populations exhibit collective dynamics that are not well captured by inspecting component agents in isolation. An individually aligned agent may participate in an emergent information cascade. An agent that would not independently deceive may become part of a coalition that collectively deceives. Recent work suggests that LLM agents can engage in tacit collusion without explicit collusive instructions, exhibit conformity under peer influence, develop conventions through decentralized interaction, and generate cascading failures through coordinated behaviour \cite{Lin2024,Cho2025,Weng2025,Ashery2025,Cemri2025}. 
 
Two responses have begun to address this problem. The first is taxonomic. Taxonomies classify multi-agent failure modes and provide a structured vocabulary for distinct risk categories \cite{Bisconti2025,Hammond2025}. The second is observational. Empirical and simulation-based studies document collective phenomena in populations of LLM agents \cite{DeMarzoGarcia2023,Park2023,Piao2025}. Both are necessary. Taxonomies identify relevant targets of inquiry. Observational studies establish that collective phenomena occur. However, neither approach, by itself, identifies which micro-level conditions are sufficient to generate those phenomena or which interventions reliably suppress them.
 
This paper proposes two linked concepts to address that gap. \textbf{Agentic microphysics} names the level of analysis: local interaction dynamics among agents, governed by the rules and affordances of their environments. \textbf{Generative safety} names the methodology: growing target phenomena from explicit micro-level conditions in order to identify sufficient conditions to produce the phenomena under investigation.
 
The distinctive feature of LLM populations is that interaction protocols are designable. Properties like synchronous vs asynchronous interaction, visibility regimes, memory access, communication formats and privilege distribution are architecture choices. If harmful collective dynamics arise from specific micro-configurations, and those configurations are design variables, then understanding microphysics creates an opportunity for architectural prevention. On this view, safety becomes a matter of protocol design of agentic environments alongside the alignment of individual models.
 
\section{From Single-Agent to Multi-Agent Risk}
 
Single-agent safety concerns are well documented, and we will limit ourselves to just hint at the most relevant literature. Learned optimizers may pursue objectives that diverge from those represented in training \cite{Hubinger2019}. Models may selectively comply with harmful requests when they infer that outputs affect training or oversight \cite{Greenblatt2024}. In-context scheming, including behaviour directed at disabling oversight, has been documented in evaluations of frontier models \cite{Meinke2024}. Agentic misalignment patterns resembling insider-threat behaviour have also been discussed in systems with tool access and persistent memory \cite{Lynch2025}. These findings concern a different explanatory problem than the one examined here.

We posit that there are two modes in which risks from agentic AI emerge. The first is \emph{emergence from agentic affordance}: safety-relevant behaviour arises from the coupling between a single agent and its environment, including tools, memory, and operational affordances. The second is \emph{emergence from multi-agent interaction}: risks arise through communication, imitation, strategic influence, and coordination among multiple agents. The explanatory target in this paper is the second case, namely agent--agent dynamics and the population-level patterns they generate.

\begin{table}[ht]
\centering
\small
\caption{Emergence from agentic affordance and emergence from multi-agent interaction.}
\label{tab:emergence}
\vspace{0.5em}
\begin{tabularx}{\textwidth}{@{}lllX@{}}
\toprule
\textbf{Level} & \textbf{Type of Emergence} & \textbf{Object of Analysis} & \textbf{Examples} \\
\midrule
Single-agent & Agentic affordance & Agent + environment/tools & Scheming, alignment faking, sycophancy \\
Micro & Multi-agent interaction & Local interaction rules & Semantic/Behavioural drifts\\
Meso & Multi-agent interaction & Group/network configurations & Coalition formation, polarization \\
Macro & Multi-agent interaction & Population-level dynamics & Cascades \\
\bottomrule
\end{tabularx}
\end{table}
 
Individually aligned agents can participate in harmful collective dynamics. A population of agents, each of which would not independently spread misinformation, may still generate an information cascade that amplifies false beliefs. Agents that avoid deceptive coalitions under isolated evaluation may form them once embedded in communication networks with particular structures. Experimental work supports this broader concern. LLM agents in Cournot competition can learn supra-competitive market-division strategies without explicit collusive instructions \cite{Lin2024}. Populations of LLM agents can develop shared conventions and collective biases through decentralized interaction \cite{Ashery2025}. Behaviour in prisoner's dilemma and public goods settings is sensitive to interaction structure and incentive design \cite{Fontana2025,Liang2025}. Multi-agent collectives also exhibit vulnerability to social influence that degrades decision quality \cite{Cho2025}. These cases indicate that safety analysis must account for population dynamics rather than only individual capabilities.
 
\section{Existing Approaches and Their Limits}
 
A first response to multi-agent risk can be considered taxonomic. In previous work \cite{Bisconti2025} we developed a taxonomy organized by micro, meso, and macro levels, distinguishing local interaction mechanisms from system-wide outcomes. On top of that, Hammond et al.\ catalogue multi-agent risks for advanced AI, identifying three broad failure modes---miscoordination, conflict, and collusion---alongside recurrent risk factors such as information asymmetries, network effects, and emergent agency \cite{Hammond2025}. 
 
A taxonomy can establish emergent phenomena like collusion, herding, or various drifts in the semantic and behavioural properties of the models. It cannot however determine which interaction rules, information structures, or agent compositions are sufficient to generate them. Moreover, It cannot determine whether a risk appears only beyond a population threshold, whether it depends on sequential rather than simultaneous decisions, or whether modest protocol changes suppress it. These are questions about \textbf{causal dynamics} and therefore require a different methodology.
 
The second response is observational. A growing literature examines collective behaviour in LLM agent populations. Existing work shows that generative agents equipped with memory and planning can produce emergent social behaviours such as coordinating events and spreading information through social networks \cite{Park2023}. This line of research has since been extended to much larger populations in order to study emergent social dynamics at scale \cite{Piao2025}. Other studies show that interaction among LLM agents can generate scale-free network structures \cite{DeMarzoGarcia2023}, reproduce opinion dynamics comparable to polarization and echo chambers familiar from bounded-confidence models \cite{Chuang2024}, and support mitigation strategies based on active and passive nudges \cite{Wang2025}. More recent work strengthens this observational picture further. Decentralized populations of LLM agents have been shown to converge on shared social conventions and to generate collective bias even when individual agents do not exhibit that bias in isolation \cite{Ashery2025}. Multi-agent social-media simulations of major public events suggest that guidance agents can reduce negative sentiment propagation and alleviate polarization \cite{Zhang2026SITGNet}. Work on networks of cognitive agents likewise indicates that information-flow structure and rapid consensus formation substantially shape emergent collective dynamics \cite{ZomerDeDomenico2026}.

Observation alone, however, cannot isolate causal mechanisms. What appears in deployed or complex simulated populations reflects a mixture of causes, including platform affordances, model homogeneity or heterogeneity, human intervention, and timing effects. Observation can establish the existence of a macro-regularity without identifying which micro-conditions are sufficient to generate it.
 
The challenge is especially acute for generative agent-based models, where validation remains the central methodological problem \cite{LarooijTornberg2026}. The black-box structure of LLMs and their stochastic outputs may intensify traditional difficulties in validating agent-based simulations. Recent work on operational validation using digital twins of online platforms illustrates both the promise and the difficulty of calibrating LLM simulations against empirical data \cite{Rossetti2025}. 

In order to integrate these two approaches, we argue that a \textbf{generative approach} is required. Such an approach links the macro-dynamics observed in agentic populations, together with the risks they produce, to the \textbf{micro-specifications} that define the mechanisms generating them. Its purpose is to move from classification and observation to explicit reconstruction: to show how population-level outcomes arise from local interaction rules, protocol conditions, and architectural constraints. We call this approach \textbf{agentic microphysics}.
 
\section{Agentic Microphysics: The Level of Analysis}
 
The term \textbf{agentic microphysics} borrows from Foucault's account of a microphysics of power, where large-scale order is reproduced through dispersed and local relations rather than exhausted by a centralized mechanism \cite{Foucault1977}. Similarly, many collective risks in multi-agent AI systems are generated through recurrent local interactions among agents, so explanation cannot stop at the level of aggregate outcomes. Agentic microphysics denotes the\textbf{ level of analysis concerned }with those interactional processes and with the population-level patterns they produce. Its aim is mechanistic, specifying the organized sequence of entities, activities, and relations through which a macro-level phenomenon is generated. The task of agentic microphysics (as a research  is to connect structural conditions, situated action, and emergent macro-outcomes through an explicit causal mechanism \cite{HedstromYlikoski2010}.
 
Safety-relevant phenomena in multi-agent systems are micro-to-macro outcomes. Cascades, collusion, polarization, strategic convention formation, and semantic drift arise through repeated episodes in which one agent's output becomes part of another agent's informational environment and decision problem. The explanatory target is therefore the generative pathway by which local exposures, updating rules, and strategic responses accumulate into comparatively stable collective patterns.  Threshold models, diffusion processes, and self-reinforcing dynamics explain how local responses generate aggregate regularities \cite{Granovetter1978,Merton1948}.
 
A commitment to agentic microphysics does not require explanatory reduction to the individual level. Mechanism-based explanation is commonly structured through a macro--micro--macro sequence. Institutional rules, network topology, incentive structures, and visibility constraints shape the situations agents face; agents act under those conditions; and the aggregate effect of those actions reproduces or transforms the macro-order. The local level remains central because it is where the operative interaction mechanism runs, but macro-level structures enter as causal inputs and macro-level outcomes remain the \textit{explananda}. An account of collective emergent phenomena within a population of agents (human and artificial) is incomplete when it identifies the initial structure and the final pattern but leaves unspecified the interaction process connecting them \cite{Ylikoski2021}.
 
At this level, the main explanatory variables are those governing inter-agent exposure, response, and adaptation. These might include  the communication protocol, namely who can address whom and in what representation; the visibility regime, namely which outputs are observable to which agents and with what latency; the turn structure, including whether decisions are sequential or simultaneous and whether ordering is fixed or endogenous; the memory regime, namely what prior interactions are retained, retrieved, or summarized; and the environmental affordances that mediate action, such as tools, APIs, scoring rules. Together these features define the \emph{interaction architecture}. That architecture belongs inside the mechanism under study because interventions on communication, ordering, or memory can alter the pathway through which local responses scale into collective outcomes. Recent surveys of LLM-based multi-agent systems identify communication, memory, and workflow design as central determinants of system behavior \cite{Luo2025,Han2024}.
 
The\textbf{ epistemic value }of agentic microphysics lies therefore in identifying explanatory relevance through controlled variation. If a collective pattern changes when visibility is perturbed, when turn order is randomized, when memory is truncated, or when population homogeneity is reduced, those features gain evidence of causal relevance. If the macro-pattern remains stable across such interventions, those features are less likely to belong to the operative mechanism. This approach therefore functions as a mechanism-discovery and mechanism-testing framework for multi-agent AI safety, linking descriptive analysis of emergent behaviour with intervention-oriented research \cite{Mahoney2012}.
 
\subsection{Adequacy Conditions for Agentic Microphysics}
 
Agentic microphysics posits that a good micro-specification of a collective phenomena satisfy three adequacy conditions: \textbf{descriptive adequacy}, \textbf{explanatory adequacy}, and \textbf{observational adequacy}.\footnote{The adequacy criteria used here are adapted from the hierarchy introduced in generative approach to the study of human syntax developed by MIT linguist Noam Chomsky \cite{Chomsky1965, Chomsky2000, Rizzi2017}. In the generative tradition, the central task is to specify a finite and explicit system of rules or principles that generates the structured expressions of a language. The aim of the \emph{agentic microphysics} approach is, similarly, to identify the smallest relevant set of interaction rules and affordances, from which such phenomena can be generated. The appeal to generative grammar provides a model of inquiry in which complex observable patterns are explained by constructing an explicit generative system and then evaluating that system in terms of what it reproduces, how accurately it specifies the underlying structure, and whether it supports explanation.}
 
\begin{enumerate}[label=(\roman*)]
\item Can the model generate the phenomenon of interest?
\item Does it identify the process that produces that phenomenon?
\item Is that process adequately supported by empirical evidence from real-world use cases?
\end{enumerate}
 
First, a microphysical account should satisfy \emph{descriptive adequacy}. A microspecified interaction model should be able to generate the target phenomenon at the level at which it is observed. In the present context, this means that explicit rules governing the interaction among agents should be sufficient to produce the relevant macro-pattern. A theory of collective behaviour that cannot generate the phenomenon under study has not yet shown that its proposed local rules are sufficient for the pattern it seeks to explain. \cite{MacyWiller2002}.
 
Second, a microphysical account should satisfy \emph{explanatory adequacy}. Reproducing an outcome is not enough if the model does so by implicitly assuming the phenomenon or by fitting the outcome without identifying the process that generates it. Explanation has force when it opens the connection between initial conditions and aggregate outcomes and specifies the entities, relations, and activities through which the outcome is produced. Agent-based models are useful in this respect because they make interaction sequences explicit and therefore permit comparison between competing candidate mechanisms \cite{Tilly2001,Elsenbroich2012}.
 
Third, a microphysical account should satisfy \emph{observational adequacy}. Even a descriptively successful and mechanistically interpretable model \textit{in silico} remains incomplete unless it is confronted with evidence of the real phenomen under scrutiny outside the model. Empirical adequacy requires that the generated pattern match the target real-world phenomenon along those dimensions that matter for the research question. These may include temporal profile, distributional shape, threshold behaviour, sensitivity to perturbation, or dependence on contextual conditions. A model that generates a plausible qualitative pattern but fails under calibration, comparative validation, or intervention does not yet support strong causal claims \cite{BruchAtwell2015,PozzoniKaidesoja2021}.
 
We mantain that microspecification is \textbf{descriptively useful} because it provides with the minimum description needed to produce the phenomenon from explicit local conditions. It is \textbf{explanatorily useful} because it identifies the interaction process that generates the phenomenon rather than relying on functionally equivalent specifications that produce the same at macro-level phenomenon. It is \textbf{observationally useful} because it creates a structure that can be calibrated, perturbed, and compared with evidence from real or experimentally controlled systems.

\section{Generative Safety: The Methodology}

If collective risks in multi-agent AI are generated through local interaction mechanisms, then identifying those mechanisms cannot rely on taxonomy alone. A further step is required: one must specify the interaction architecture in sufficiently explicit terms that the phenomenon can be reconstructed from it, varied under controlled conditions, and compared against observational evidence.

\textbf{Generative safety} names this methodological step. It provides the experimental logic through which agentic microphysics becomes an explanatory and intervention-oriented research program rather than only a level of description.
The methodological template we adopt comes from generative social science. Epstein and Axtell showed that macro-level regularities such as group formation, cultural transmission, and conflict could be grown from local interaction rules among decentralized agents \cite{EpsteinAxtell1996}. Epstein later formulated the associated epistemic standard: if a phenomenon has not been grown \emph{in silico}, its emergence has not been explained \cite{Epstein1999,Epstein2006}. Schelling's segregation model showed that mild local preferences can generate strong residential segregation \cite{Schelling1971}. Axelrod's tournaments showed that stable cooperation can emerge from simple repeated-game strategies without central enforcement \cite{Axelrod1984}.
 
Two methodological commitments structure the generative approach. The first is a mechanism-based form of \emph{methodological individualism}: population-level outcomes are explained through the situations agents face, the local rules they follow, and the aggregate consequences of their actions \cite{Weber1922,Coleman1990}. The second (as we saw in the last section) is \emph{microspecification}: local interaction rules must be stated precisely enough that the model's dynamics are determined by those rules.
 
\textbf{Generative safety} proceeds through five stages.
 
\medskip
\noindent\textit{Stage 1: Risk identification.} The starting point is a structured taxonomy of micro, meso or macro-level phenomena constituting safety targets, such as collusion, information cascades, manipulation, coordination failure, polarization, and deception.
 
\medskip
\noindent\textit{Stage 2: Microspecification.} For each target risk, formulate a concrete hypothesis about local interaction rules and environmental conditions sufficient to generate it. 
 
\medskip
\noindent\textit{Stage 3: Generative experimentation.} Implement the hypothesized micro-configuration in a controlled multi-agent environment. The central question is whether the target phenomenon emerges. Parameter variation identifies sufficient conditions, thresholds, and sensitivity.
 
\medskip
\noindent\textit{Stage 4: Intervention design.} Once a mechanism has been identified, the same environment can be used to test interventions. Interventions may target the model or the interaction architecture. Recent work on governing LLM collusion illustrates this distinction empirically: prompt-only prohibitions may fail to suppress collusive outcomes under incentive pressure, whereas externalized governance mechanisms with runtime enforcement can reduce harmful coordination \cite{Bracale2026}. Related game-theoretic work shows that protocol structure alone can shift equilibrium behaviour in sequential public goods settings \cite{Liang2025}.
 
\medskip
\noindent\textit{Stage 5: Observational validation.} Compare results from generative experiments against "field" data from deployed settings. Agreement supports external validity. Divergence indicates incomplete microspecification, poor calibration or confounded observation. Operational validation studies demonstrate this logic by comparing simulated platform dynamics against empirical baselines across activity patterns, network structure, and content distributions \cite{Rossetti2025}.
 
\begin{figure}[ht]
  \centering
  \includegraphics[width=0.65\textwidth]{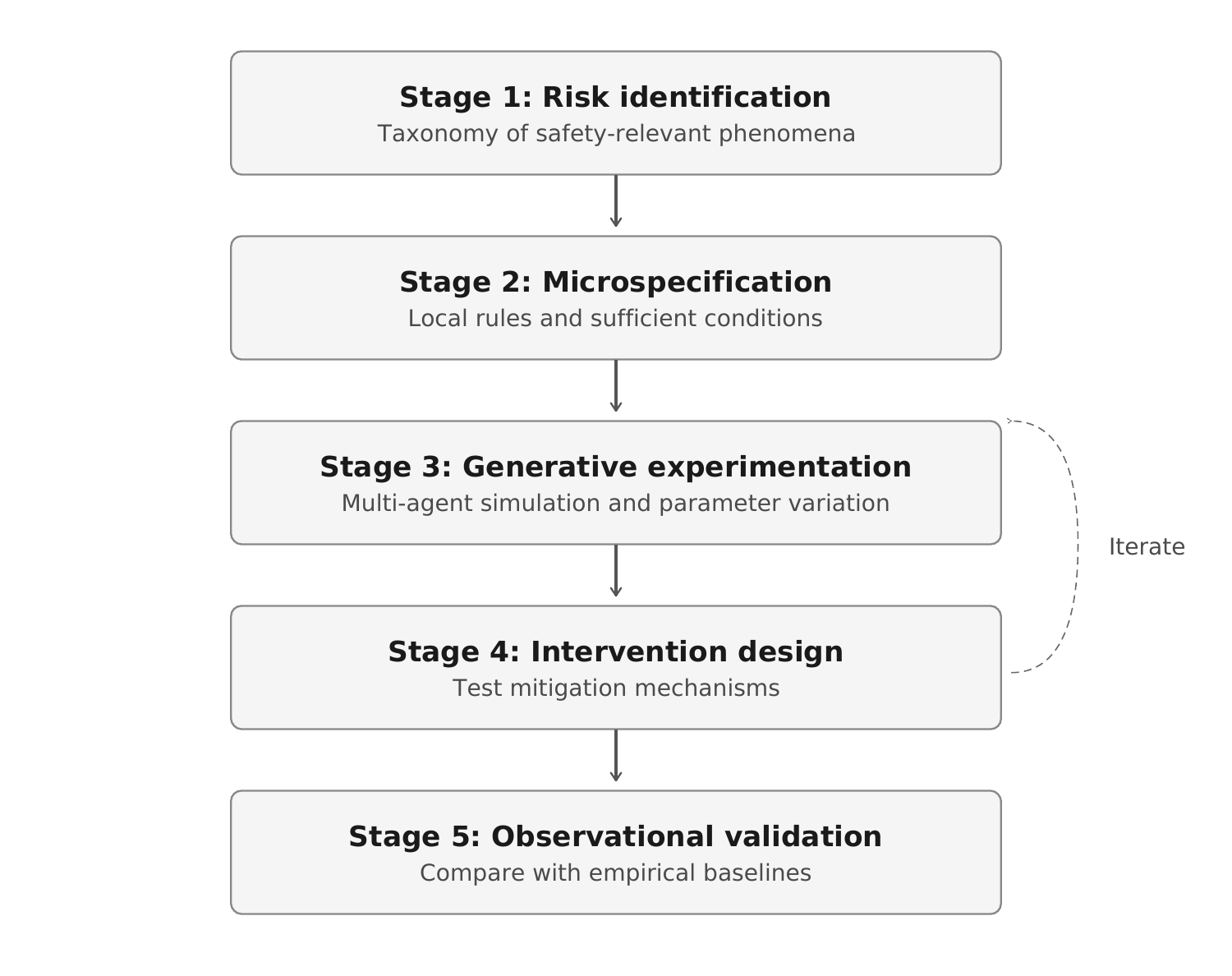}
  \caption{The generative safety pipeline. \textbf{Stage~1} identifies macro-level risk phenomena (e.g., collusion, polarization). \textbf{Stage~2} formulates testable hypotheses about local interaction rules sufficient to generate them. \textbf{Stage~3} implements these configurations in controlled multi-agent simulations. \textbf{Stage~4} uses the same environment to test interventions targeting model behavior or interaction architecture. \textbf{Stage~5} validates results against empirical baselines. The dashed arrow indicates iteration between experimentation and intervention design.}
  \label{fig:generative-safety}
\end{figure}

\section{Applied Microphysics: Herding in LLM News-Feed Environments}
 
We applied the generative safety methodology to a multi agent controlled setting \cite{Prandi2026Herding}. We simulated a minimal social-media environment (structured on Moltbook) in which LLM agents interact through a shared news feed, allowing direct manipulation of micro-level interaction variables.  We investigate the emergence of herding among AI agents. 
We use \emph{herding} to denote a collective pattern in which agents converge on the same items or choices given the behavior of other agents. The experiment isolates two candidate drivers of herding: visible social proof (through the amount of likes the news have in the feed) and presentation order (their position in the feed the agents interact with). The objective is to identify which local mechanisms have causal efficacy in shaping collective attention.
 
The design follows the logic of agentic microphysics. The interaction is reduced to its minimal components: a fixed content slate, shuffled presentation order, visible or hidden engagement signals, and stateless endorsement decisions. This enables clean identification of which variables govern the transition from individual evaluation to collective convergence.
 
\paragraph{Environment and microspecification}
Agents repeatedly browse a feed of 48 items under equal exposure conditions. On each interaction, the ordering of items is randomly reshuffled, and agents decide which items to endorse. The experiment varies the visibility of prior endorsements across conditions, including hidden signals, organically accumulated signals, and seeded popularity levels.

\paragraph{Results}
The central empirical result is that feed position, rather than visible popularity, governs collective attention. Agents select almost exclusively from the top-ranked items, with mean selected positions concentrated around the first few slots in a 48-item feed. Visible social proof shifts behaviour only within this restricted choice set and exhibits a threshold effect: the presence of any positive signal increases selection probability, but increasing the magnitude of the signal produces no systematic additional effect.
 
Low-ranked items remain effectively unselected even when associated with strong visible endorsement. This indicates that social proof cannot compensate for positional disadvantage in this environment. The collective pattern is therefore generated by a two-stage mechanism: positional gating determines the effective choice set, and social proof modulates selection within that set.

From a \textbf{safety perspective}, these findings identify a specific causal structure underlying herding in LLM populations that can be used as a vector of attack by a malicious actor. Feed ranking determines which content enters the agents' effective decision space in the first place.  An \textbf{attacker} could exploit this mechanism by manipulating the ranking process so that selected items appear systematically in the first visible positions. Because positional exposure governs the effective choice set, such manipulation can induce disproportionate collective attention even without highly persuasive content or large endorsement counts.  Ranking-based perturbations could therefore induce synchronized shifts of attention across otherwise heterogeneous agents, creating a common architectural failure mode.  This phenomena might appear as a form of collective drift \textbf{spontaneously emerging} from repeated interactions among agents.

The application we described illustrates the role of agentic microphysics (and generative safety) within the broader research programme. A relatively small, controlled experiment is sufficient to identify a causal mechanism that would remain underdetermined in larger and more complex simulations. Designers of agentic environments can leverage this knowledge to define \textbf{measurable interventions} against multi-agent risks, thereby increasing the resilience and robustness of agentic ecosystems.

\end{document}